\documentclass[conference,10pt]{IEEEtran}
\usepackage[utf8]{inputenc}
\ifCLASSINFOpdf
  \usepackage[pdftex]{graphicx}
\else
  \usepackage[dvips]{graphicx}
\fi
\usepackage{algpseudocode}
\usepackage{amsmath}
\usepackage{color}
\usepackage{doi}
\usepackage{enumitem}
\usepackage{etoolbox}
\usepackage{fancyvrb}
\usepackage{float}
\usepackage{hyperref}
\usepackage{multicol}
\usepackage{pbox}
\usepackage{romannum}
\usepackage{tabularx}

\makeatletter

\hypersetup{
    colorlinks,
    linkcolor={black},
    citecolor={black},
    urlcolor={blue}
}
\patchcmd{\@maketitle}
  {\addvspace{0.5\baselineskip}\egroup}
  {\addvspace{-1.5\baselineskip}\egroup}
  {}
  {}
\def\PY@reset{\let\PY@it=\relax \let\PY@bf=\relax%
    \let\PY@ul=\relax \let\PY@tc=\relax%
    \let\PY@bc=\relax \let\PY@ff=\relax}
\def\PY@tok#1{\csname PY@tok@#1\endcsname}
\def\PY@toks#1+{\ifx\relax#1\empty\else%
    \PY@tok{#1}\expandafter\PY@toks\fi}
\def\PY@do#1{\PY@bc{\PY@tc{\PY@ul{%
    \PY@it{\PY@bf{\PY@ff{#1}}}}}}}
\def\PY#1#2{\PY@reset\PY@toks#1+\relax+\PY@do{#2}}

\expandafter\def\csname PY@tok@gd\endcsname{\def\PY@tc##1{\textcolor[rgb]{0.63,0.00,0.00}{##1}}}
\expandafter\def\csname PY@tok@gu\endcsname{\let\PY@bf=\textbf\def\PY@tc##1{\textcolor[rgb]{0.50,0.00,0.50}{##1}}}
\expandafter\def\csname PY@tok@gt\endcsname{\def\PY@tc##1{\textcolor[rgb]{0.00,0.27,0.87}{##1}}}
\expandafter\def\csname PY@tok@gs\endcsname{\let\PY@bf=\textbf}
\expandafter\def\csname PY@tok@gr\endcsname{\def\PY@tc##1{\textcolor[rgb]{1.00,0.00,0.00}{##1}}}
\expandafter\def\csname PY@tok@cm\endcsname{\let\PY@it=\textit\def\PY@tc##1{\textcolor[rgb]{0.25,0.50,0.50}{##1}}}
\expandafter\def\csname PY@tok@vg\endcsname{\def\PY@tc##1{\textcolor[rgb]{0.10,0.09,0.49}{##1}}}
\expandafter\def\csname PY@tok@vi\endcsname{\def\PY@tc##1{\textcolor[rgb]{0.10,0.09,0.49}{##1}}}
\expandafter\def\csname PY@tok@mh\endcsname{\def\PY@tc##1{\textcolor[rgb]{0.40,0.40,0.40}{##1}}}
\expandafter\def\csname PY@tok@cs\endcsname{\let\PY@it=\textit\def\PY@tc##1{\textcolor[rgb]{0.25,0.50,0.50}{##1}}}
\expandafter\def\csname PY@tok@ge\endcsname{\let\PY@it=\textit}
\expandafter\def\csname PY@tok@vc\endcsname{\def\PY@tc##1{\textcolor[rgb]{0.10,0.09,0.49}{##1}}}
\expandafter\def\csname PY@tok@il\endcsname{\def\PY@tc##1{\textcolor[rgb]{0.40,0.40,0.40}{##1}}}
\expandafter\def\csname PY@tok@go\endcsname{\def\PY@tc##1{\textcolor[rgb]{0.53,0.53,0.53}{##1}}}
\expandafter\def\csname PY@tok@cp\endcsname{\def\PY@tc##1{\textcolor[rgb]{0.74,0.48,0.00}{##1}}}
\expandafter\def\csname PY@tok@gi\endcsname{\def\PY@tc##1{\textcolor[rgb]{0.00,0.63,0.00}{##1}}}
\expandafter\def\csname PY@tok@gh\endcsname{\let\PY@bf=\textbf\def\PY@tc##1{\textcolor[rgb]{0.00,0.00,0.50}{##1}}}
\expandafter\def\csname PY@tok@ni\endcsname{\let\PY@bf=\textbf\def\PY@tc##1{\textcolor[rgb]{0.60,0.60,0.60}{##1}}}
\expandafter\def\csname PY@tok@nl\endcsname{\def\PY@tc##1{\textcolor[rgb]{0.63,0.63,0.00}{##1}}}
\expandafter\def\csname PY@tok@nn\endcsname{\let\PY@bf=\textbf\def\PY@tc##1{\textcolor[rgb]{0.00,0.00,1.00}{##1}}}
\expandafter\def\csname PY@tok@no\endcsname{\def\PY@tc##1{\textcolor[rgb]{0.53,0.00,0.00}{##1}}}
\expandafter\def\csname PY@tok@na\endcsname{\def\PY@tc##1{\textcolor[rgb]{0.49,0.56,0.16}{##1}}}
\expandafter\def\csname PY@tok@nb\endcsname{\def\PY@tc##1{\textcolor[rgb]{0.00,0.50,0.00}{##1}}}
\expandafter\def\csname PY@tok@nc\endcsname{\let\PY@bf=\textbf\def\PY@tc##1{\textcolor[rgb]{0.00,0.00,1.00}{##1}}}
\expandafter\def\csname PY@tok@nd\endcsname{\def\PY@tc##1{\textcolor[rgb]{0.67,0.13,1.00}{##1}}}
\expandafter\def\csname PY@tok@ne\endcsname{\let\PY@bf=\textbf\def\PY@tc##1{\textcolor[rgb]{0.82,0.25,0.23}{##1}}}
\expandafter\def\csname PY@tok@nf\endcsname{\def\PY@tc##1{\textcolor[rgb]{0.00,0.00,1.00}{##1}}}
\expandafter\def\csname PY@tok@si\endcsname{\let\PY@bf=\textbf\def\PY@tc##1{\textcolor[rgb]{0.73,0.40,0.53}{##1}}}
\expandafter\def\csname PY@tok@s2\endcsname{\def\PY@tc##1{\textcolor[rgb]{0.73,0.13,0.13}{##1}}}
\expandafter\def\csname PY@tok@nt\endcsname{\let\PY@bf=\textbf\def\PY@tc##1{\textcolor[rgb]{0.00,0.50,0.00}{##1}}}
\expandafter\def\csname PY@tok@nv\endcsname{\def\PY@tc##1{\textcolor[rgb]{0.10,0.09,0.49}{##1}}}
\expandafter\def\csname PY@tok@s1\endcsname{\def\PY@tc##1{\textcolor[rgb]{0.73,0.13,0.13}{##1}}}
\expandafter\def\csname PY@tok@ch\endcsname{\let\PY@it=\textit\def\PY@tc##1{\textcolor[rgb]{0.25,0.50,0.50}{##1}}}
\expandafter\def\csname PY@tok@m\endcsname{\def\PY@tc##1{\textcolor[rgb]{0.40,0.40,0.40}{##1}}}
\expandafter\def\csname PY@tok@gp\endcsname{\let\PY@bf=\textbf\def\PY@tc##1{\textcolor[rgb]{0.00,0.00,0.50}{##1}}}
\expandafter\def\csname PY@tok@sh\endcsname{\def\PY@tc##1{\textcolor[rgb]{0.73,0.13,0.13}{##1}}}
\expandafter\def\csname PY@tok@ow\endcsname{\let\PY@bf=\textbf\def\PY@tc##1{\textcolor[rgb]{0.67,0.13,1.00}{##1}}}
\expandafter\def\csname PY@tok@sx\endcsname{\def\PY@tc##1{\textcolor[rgb]{0.00,0.50,0.00}{##1}}}
\expandafter\def\csname PY@tok@bp\endcsname{\def\PY@tc##1{\textcolor[rgb]{0.00,0.50,0.00}{##1}}}
\expandafter\def\csname PY@tok@c1\endcsname{\let\PY@it=\textit\def\PY@tc##1{\textcolor[rgb]{0.25,0.50,0.50}{##1}}}
\expandafter\def\csname PY@tok@o\endcsname{\def\PY@tc##1{\textcolor[rgb]{0.40,0.40,0.40}{##1}}}
\expandafter\def\csname PY@tok@kc\endcsname{\let\PY@bf=\textbf\def\PY@tc##1{\textcolor[rgb]{0.00,0.50,0.00}{##1}}}
\expandafter\def\csname PY@tok@c\endcsname{\let\PY@it=\textit\def\PY@tc##1{\textcolor[rgb]{0.25,0.50,0.50}{##1}}}
\expandafter\def\csname PY@tok@mf\endcsname{\def\PY@tc##1{\textcolor[rgb]{0.40,0.40,0.40}{##1}}}
\expandafter\def\csname PY@tok@err\endcsname{\def\PY@bc##1{\setlength{\fboxsep}{0pt}\fcolorbox[rgb]{1.00,0.00,0.00}{1,1,1}{\strut ##1}}}
\expandafter\def\csname PY@tok@mb\endcsname{\def\PY@tc##1{\textcolor[rgb]{0.40,0.40,0.40}{##1}}}
\expandafter\def\csname PY@tok@ss\endcsname{\def\PY@tc##1{\textcolor[rgb]{0.10,0.09,0.49}{##1}}}
\expandafter\def\csname PY@tok@sr\endcsname{\def\PY@tc##1{\textcolor[rgb]{0.73,0.40,0.53}{##1}}}
\expandafter\def\csname PY@tok@mo\endcsname{\def\PY@tc##1{\textcolor[rgb]{0.40,0.40,0.40}{##1}}}
\expandafter\def\csname PY@tok@kd\endcsname{\let\PY@bf=\textbf\def\PY@tc##1{\textcolor[rgb]{0.00,0.50,0.00}{##1}}}
\expandafter\def\csname PY@tok@mi\endcsname{\def\PY@tc##1{\textcolor[rgb]{0.40,0.40,0.40}{##1}}}
\expandafter\def\csname PY@tok@kn\endcsname{\let\PY@bf=\textbf\def\PY@tc##1{\textcolor[rgb]{0.00,0.50,0.00}{##1}}}
\expandafter\def\csname PY@tok@cpf\endcsname{\let\PY@it=\textit\def\PY@tc##1{\textcolor[rgb]{0.25,0.50,0.50}{##1}}}
\expandafter\def\csname PY@tok@kr\endcsname{\let\PY@bf=\textbf\def\PY@tc##1{\textcolor[rgb]{0.00,0.50,0.00}{##1}}}
\expandafter\def\csname PY@tok@s\endcsname{\def\PY@tc##1{\textcolor[rgb]{0.73,0.13,0.13}{##1}}}
\expandafter\def\csname PY@tok@kp\endcsname{\def\PY@tc##1{\textcolor[rgb]{0.00,0.50,0.00}{##1}}}
\expandafter\def\csname PY@tok@w\endcsname{\def\PY@tc##1{\textcolor[rgb]{0.73,0.73,0.73}{##1}}}
\expandafter\def\csname PY@tok@kt\endcsname{\def\PY@tc##1{\textcolor[rgb]{0.69,0.00,0.25}{##1}}}
\expandafter\def\csname PY@tok@sc\endcsname{\def\PY@tc##1{\textcolor[rgb]{0.73,0.13,0.13}{##1}}}
\expandafter\def\csname PY@tok@sb\endcsname{\def\PY@tc##1{\textcolor[rgb]{0.73,0.13,0.13}{##1}}}
\expandafter\def\csname PY@tok@k\endcsname{\let\PY@bf=\textbf\def\PY@tc##1{\textcolor[rgb]{0.00,0.50,0.00}{##1}}}
\expandafter\def\csname PY@tok@se\endcsname{\let\PY@bf=\textbf\def\PY@tc##1{\textcolor[rgb]{0.73,0.40,0.13}{##1}}}
\expandafter\def\csname PY@tok@sd\endcsname{\let\PY@it=\textit\def\PY@tc##1{\textcolor[rgb]{0.73,0.13,0.13}{##1}}}

\def\PYZus{\char`\_}

\def\PYZca{\char`\^}

\def\PYZlt{\char`\<}
\def\PYZgt{\char`\>}

\def\PYZhy{\char`\-}

\def\PYZdq{\char`\"}

\makeatother

\author{
\IEEEauthorblockN{Vadim Markovtsev\\\texttt{vadim@sourced.tech}}
\and
\IEEEauthorblockN{Eiso Kant\\\texttt{eiso@sourced.tech}}
}

\title{Topic modeling of public repositories at scale using names in source code}

\begin{document}

\twocolumn[
\begin{@twocolumnfalse}
\maketitle
\begin{center}
source\{d\}, Madrid, Spain\\February 2017
\end{center}
\end{@twocolumnfalse}
\bigskip]

\begin{abstract}
Programming languages themselves have a limited number of reserved keywords and character based tokens that define the language specification. However, programmers have a rich use of natural language within their code through comments, text literals and naming entities. The programmer defined names that can be found in source code are a rich source of information to build a high level understanding of the project. The goal of this paper is to apply topic modeling to names used in over 13.6 million repositories and perceive the inferred topics. One of the problems in such a study is the occurrence of duplicate repositories not officially marked as forks (obscure forks). We show how to address it using the same identifiers which are extracted for topic modeling.

We open with a discussion on naming in source code, we then elaborate on our approach to remove exact duplicate and fuzzy duplicate repositories using Locality Sensitive Hashing on the bag-of-words model and then discuss our work on topic modeling; and finally present the results from our data analysis together with open-access to the source code, tools and datasets.
\end{abstract}

\begin{IEEEkeywords}
programming, open source, source code, software repositories, git, GitHub, topic modeling, ARTM, locality sensitive hashing, MinHash, open dataset, data.world.
\end{IEEEkeywords}

\section{Introduction}

\IEEEPARstart{T}{here} are more than 18 million non-empty public repositories on \href{https://github.com}{GitHub} which are not marked as forks. This makes GitHub the largest version control repository hosting service. It has become difficult to explore such a large number of projects and nearly impossible to classify them. One of the main sources of information that exists about public repositories is their code.

To gain a deeper understanding of software development it is important to understand the trends among open-source projects. Bleeding edge technologies are often used first in open source projects and later employed in proprietary solutions when they become stable enough\footnote{Notable examples include the Linux OS kernel, the PostgreSQL database engine, the Apache Spark cluster-computing framework and the Docker containers.}. An exploratory analysis of open-source projects can help to detect such trends and provide valuable insight for industry and academia. 

Since GitHub appeared the open-source movement has gained significant momentum. Historically developers would manually register their open-source projects in software digests. As the number of projects dramatically grew, those lists became very hard to update; as a result they became more fragmented and started exclusively specializing in narrow technological ecosystems. The next attempt to classify open source projects was based on manually submitted lists of keywords. While this approach works \cite{sourceforge}, it requires careful keywords engineering to appear comprehensive, and thus not widely adopted by the end users in practice. GitHub introduced repository tags in January 2017 which is a variant of manual keywords submission.

The present paper describes how to conduct fully automated topic extraction from millions of public repositories. It scales linearly with the overall source code size and has substantial performance reserve to support the future growth. We propose building a bag-of-words model on names occurring in  source code and applying proven Natural Language Processing algorithms to it. Particularly, we describe how "Weighted MinHash" algorithm \cite{Ioffe:2010:ICS:1933307.1934593} helps to filter fuzzy duplicates and how an Additive Regularized Topic Model (ARTM) \cite{Vorontsov2015} can be efficiently trained. The result of the topic modeling is a nearly-complete open source projects classification. It reflects the drastic variety in open source projects and reflect multiple features. The dataset we work with consists of approx. 18 million public repositories retrieved from GitHub in October 2016.

The rest of the paper is organised as follows: Section \Romannum{2} reviews prior work on the subject. Section \Romannum{3} elaborates on how we turn software repositories into bags-of-words. Section \Romannum{4} describes the approach to efficient filtering of fuzzy repository clones. Section \Romannum{5} covers the building of the ARTM model with 256 manually labeled topics. Section \Romannum{6} presents the achieved topic modeling results. Section \Romannum{7} lists the opened datasets we were able to prepare. Finally, section \Romannum{8} presents a conclusion and suggests improvements to future work.

\section{Related work}
\subsection{Academia}
There was an open source community study which presented statistics about manually picked topics in 2005 by J. Xu et.al. \cite{JinXu2005}.

Blincoe et.al. \cite{Blincoe:2015:EGM:2820518.2820544} studied GitHub ecosystems using reference coupling over the GHTorrent dataset \cite{Gousi13} which contained 2.4 million projects. This research employs an alternative topic modeling method on source code of 13.6 million projects.

Instead of using the GHTorrent dataset we've prepared open datasets from almost all public repositories on GitHub to be able to have a more comprehensive overview. 

M. Lungi \cite{Lungu} conducted an in-depth study of software ecosystems in 2009, the year when GitHub appeared. The examples in this work used samples of approx. 10 repositories. And the proposed discovery methods did not include Natural Language Processing.

The problem of the correct analysis of forks on GitHub has been discussed by Kalliamvakou et.al. \cite{Kalliamvakou:2014:PPM:2597073.2597074} along with other valuable concerns.

Topic modeling of source code has been applied to a variety of problems reviewed in \cite{7515925}: improvement of software maintenance \cite{6178887}, \cite{6747182}, defects explanation \cite{Chen:2012:ESD:2664446.2664476}, concept analysis \cite{4656403}, \cite{Linstead:2007:MCC:1321631.1321709}, software evolution analysis \cite{4725072}, \cite{Thomas:2011:MET:1985441.1985467}, finding similarities and clones \cite{889845}, clustering source code and discovering the internal structure \cite{802296}, \cite{Kuhn:2007:SCI:1224560.1224698}, \cite{Thomas:2011:MSR:1985793.1986020}, summarizing \cite{6613829}, \cite{McBurney:2014:ITM:2597008.2597793}, \cite{Saeidi:2015:IIT:2820282.2820331}. In the aforementioned works, the scope of the research was focused on individual projects. 

The usage of topic modeling \cite{Sun:2015:MSR:2799128.2799145} focused on improving software maintenance and was evaluated on 4 software projects. Concepts were extracted using a corpus of 24 projects in \cite{10.1109/ICPC.2010.12}. Giriprasad Sridhara et.al. \cite{10.1109/ICPC.2008.18}, Yang and Tan \cite{Yang:2012:ISR:2664446.2664472}, Howard et.al. \cite{Howard:2013:AMS:2487085.2487155} considered comments and/or names to find semantically similar terms; Haiduc and Marcus \cite{10.1109/ICPC.2008.29} researched common domain terms appearing in source code. The presented approach in this papers reveals similar and domain terms, but leverages a much significantly larger dataset of 13.6 million repositories.

Bajracharya and Lopes \cite{Bajracharya:2009:MST:1590955.1591146} trained a topic model on the year long usage log of Koders, one of the major commercial code search engines. The topic categories suggested by Bajracharya and Lopes share little similarity with the categories described in this paper since the input domain is much different.

\subsection{Industry}
To our knowledge, there are few companies which maintain a complete mirror of GitHub repositories. source\{d\} \cite{sourced} is focused on doing machine learning on top of the collected source code. Software Heritage \cite{SoftwareHeritage} strives to become web.archive.org for open source software. SourceGraph \cite{SourceGraph} processes source code references, internal and external, and created a complete reference graph for projects written in Golang.

libraries.io \cite{libraries.io} is not GitHub centric but rather processes the dependencies and metadata of open source packages fetched from a number of repositories. It analyses the dependency graph (at the level of projects, while SourceGraph analyses at the level of functions).

\section{Building the bag-of-words model} \label{bag_of_words}

This section follows the way we convert software repositories into bags-of-words, the model which stores each project as a multiset of its identifiers, ignoring the order while maintaining multiplicity. 

For the purpose of our analysis we choose to use the latest version of the master branch of each repository. And treat each repository as a single document. An improvement for further research would be to use the entire history of each repository including unique code found in each branch.

\subsection{Preliminary Processing}

Our first goal is to process each repository to identify which files contain source code, and which files are redundant for our purpose. GitHub has an open-source machine learning based library named \texttt{linguist} \cite{linguist} that identifies the programming language used within a file based on its extension and contents. We modified it to also identify vendor code and automatically generated files. The first step in our pre-processing is to run \texttt{linguist} over each repository's master branch. From 11.94 million repositories we end up with 402.6 million source files in which we have high confidence it is source code written by a developer in that project. Identifying the programming language used within each file is important for the next step, the names extraction, as it determines the programming language parser.

\subsection{Extracting Names}

Source code highlighting is a typical task for professional text editors and IDE's. There have been several open source libraries created to tackle this task. Each works by having a grammar file written per programming language which contains the rules. Pygments \cite{Pygments} is a high quality community-driven package for Python which supports more than 400 programming languages and markups. According to Pygments, all source code tokens are classified across the following categories: comments, escapes, indentations and generic symbols, reserved keywords, literals, operators, punctuation and names.

Linguist and Pygments have different sets of supported languages. Linguist stores it's list at \href{https://github.com/github/linguist/blob/master/lib/linguist/languages.yml}{master/lib/linguist/languages.yml} and the similar Pygments list is stored as \texttt{pygments.lexers.LEXERS}. Each has nearly 400 items and the intersection is approximately 200 programming languages ("programming" Linguist's item type). The languages common to Linguist and Pygments which were chosen are listed in appendix \ref{languages}. In this research we apply Pygments to the 402.6 million source files to extract all tokens which belong to the type \texttt{Token.Name}. 

\subsection{Processing names}

The next step is to process the names according to naming conventions. As an example \texttt{class FooBarBaz} adds three words to the bag: \texttt{foo}, \texttt{bar} and \texttt{baz}, or \texttt{int wdSize} should add two: \texttt{wdsize} and \texttt{size}. Fig. \ref{split_code} is the full listing of the function written in Python 3.4+ which splits identifiers.

\begin{figure}
\caption{Identifier splitting algorithm, Python 3.4+}
\label{split_code}
\begin{Verbatim}[commandchars=\\\{\},frame=single,fontsize=\small]
\PY{n}{NAME\PYZus{}BREAKUP\PYZus{}RE} \PY{o}{=} \PY{n}{re}\PY{o}{.}\PY{n}{compile}\PY{p}{(}\PY{l+s+s2}{r\PYZdq{}}\PY{l+s+s2}{[\PYZca{}a\PYZhy{}zA\PYZhy{}Z]+}\PY{l+s+s2}{\PYZdq{}}\PY{p}{)}

\PY{k}{def} \PY{n+nf}{extract\PYZus{}names}\PY{p}{(}\PY{n}{token}\PY{p}{)}\PY{p}{:}
  \PY{n}{token} \PY{o}{=} \PY{n}{token}\PY{o}{.}\PY{n}{strip}\PY{p}{(}\PY{p}{)}
  \PY{n}{prev\PYZus{}p} \PY{o}{=} \PY{p}{[}\PY{l+s+s2}{\PYZdq{}}\PY{l+s+s2}{\PYZdq{}}\PY{p}{]}

  \PY{k}{def} \PY{n+nf}{ret}\PY{p}{(}\PY{n}{name}\PY{p}{)}\PY{p}{:}
    \PY{n}{r} \PY{o}{=} \PY{n}{name}\PY{o}{.}\PY{n}{lower}\PY{p}{(}\PY{p}{)}
    \PY{k}{if} \PY{n+nb}{len}\PY{p}{(}\PY{n}{name}\PY{p}{)} \PY{o}{\PYZgt{}}\PY{o}{=} \PY{l+m+mi}{3}\PY{p}{:}
      \PY{k}{yield} \PY{n}{r}
      \PY{k}{if} \PY{n}{prev\PYZus{}p}\PY{p}{[}\PY{l+m+mi}{0}\PY{p}{]}\PY{p}{:}
        \PY{k}{yield} \PY{n}{prev\PYZus{}p}\PY{p}{[}\PY{l+m+mi}{0}\PY{p}{]} \PY{o}{+} \PY{n}{r}
        \PY{n}{prev\PYZus{}p}\PY{p}{[}\PY{l+m+mi}{0}\PY{p}{]} \PY{o}{=} \PY{l+s+s2}{\PYZdq{}}\PY{l+s+s2}{\PYZdq{}}
    \PY{k}{else}\PY{p}{:}
      \PY{n}{prev\PYZus{}p}\PY{p}{[}\PY{l+m+mi}{0}\PY{p}{]} \PY{o}{=} \PY{n}{r}

  \PY{k}{for} \PY{n}{part} \PY{o+ow}{in} \PY{n}{NAME\PYZus{}BREAKUP\PYZus{}RE}\PY{o}{.}\PY{n}{split}\PY{p}{(}\PY{n}{token}\PY{p}{)}\PY{p}{:}
    \PY{k}{if} \PY{o+ow}{not} \PY{n}{part}\PY{p}{:}
      \PY{k}{continue}
    \PY{n}{prev} \PY{o}{=} \PY{n}{part}\PY{p}{[}\PY{l+m+mi}{0}\PY{p}{]}
    \PY{n}{pos} \PY{o}{=} \PY{l+m+mi}{0}
    \PY{k}{for} \PY{n}{i} \PY{o+ow}{in} \PY{n+nb}{range}\PY{p}{(}\PY{l+m+mi}{1}\PY{p}{,} \PY{n+nb}{len}\PY{p}{(}\PY{n}{part}\PY{p}{)}\PY{p}{)}\PY{p}{:}
      \PY{n}{this} \PY{o}{=} \PY{n}{part}\PY{p}{[}\PY{n}{i}\PY{p}{]}
      \PY{k}{if} \PY{n}{prev}\PY{o}{.}\PY{n}{islower}\PY{p}{(}\PY{p}{)} \PY{o+ow}{and} \PY{n}{this}\PY{o}{.}\PY{n}{isupper}\PY{p}{(}\PY{p}{)}\PY{p}{:}
        \PY{k}{yield from} \PY{n}{ret}\PY{p}{(}\PY{n}{part}\PY{p}{[}\PY{n}{pos}\PY{p}{:}\PY{n}{i}\PY{p}{]}\PY{p}{)}
        \PY{n}{pos} \PY{o}{=} \PY{n}{i}
      \PY{k}{elif} \PY{n}{prev}\PY{o}{.}\PY{n}{isupper}\PY{p}{(}\PY{p}{)} \PY{o+ow}{and} \PY{n}{this}\PY{o}{.}\PY{n}{islower}\PY{p}{(}\PY{p}{)}\PY{p}{:}
        \PY{k}{if} \PY{l+m+mi}{0} \PY{o}{\PYZlt{}} \PY{n}{i} \PY{o}{\PYZhy{}} \PY{l+m+mi}{1} \PY{o}{\PYZhy{}} \PY{n}{pos} \PY{o}{\PYZlt{}}\PY{o}{=} \PY{l+m+mi}{3}\PY{p}{:}
          \PY{k}{yield from} \PY{n}{ret}\PY{p}{(}\PY{n}{part}\PY{p}{[}\PY{n}{pos}\PY{p}{:}\PY{n}{i} \PY{o}{\PYZhy{}} \PY{l+m+mi}{1}\PY{p}{]}\PY{p}{)}
          \PY{n}{pos} \PY{o}{=} \PY{n}{i} \PY{o}{\PYZhy{}} \PY{l+m+mi}{1}
        \PY{k}{elif} \PY{n}{i} \PY{o}{\PYZhy{}} \PY{l+m+mi}{1} \PY{o}{\PYZgt{}} \PY{n}{pos}\PY{p}{:}
          \PY{k}{yield from} \PY{n}{ret}\PY{p}{(}\PY{n}{part}\PY{p}{[}\PY{n}{pos}\PY{p}{:}\PY{n}{i}\PY{p}{]}\PY{p}{)}
          \PY{n}{pos} \PY{o}{=} \PY{n}{i}
      \PY{n}{prev} \PY{o}{=} \PY{n}{this}
    \PY{n}{last} \PY{o}{=} \PY{n}{part}\PY{p}{[}\PY{n}{pos}\PY{p}{:}\PY{p}{]}
    \PY{k}{if} \PY{n}{last}\PY{p}{:}
      \PY{k}{yield from} \PY{n}{ret}\PY{p}{(}\PY{n}{last}\PY{p}{)}
\end{Verbatim}
\end{figure}
In this step each repository is saved as an sqlite database file which contains a table with: programming language, name extracted and frequency of occurance in that repository. The total number of unique names that were extracted were  17.28 million.

\subsection{Stemming names}

It is common to stem names when creating a bag-of-words in NLP. Since we are working with natural language that is predominantly English we have applied the Snowball stemmer \cite{citeulike:1132794} from the Natural Language Toolkit (NLTK) \cite{nltk}. The stemmer was applied to names which were $>$6 characters long. In further research a diligent step would be to compare results with and without stemming of the names, and also to predetermine the language of the name (when available) and apply stemmers in different languages.

The length of words on which stemming was applied was chosen after the manual observation that shorter identifiers tend to collide with each other when stemmed and longer identifiers need to be normalized. Fig. \ref{name_lengths} represents the distribution of identifier lengths in the dataset:

\begin{figure}
\caption{Name lengths distribution}
\label{name_lengths}
\begin{center}
\includegraphics[width=0.5\textwidth]{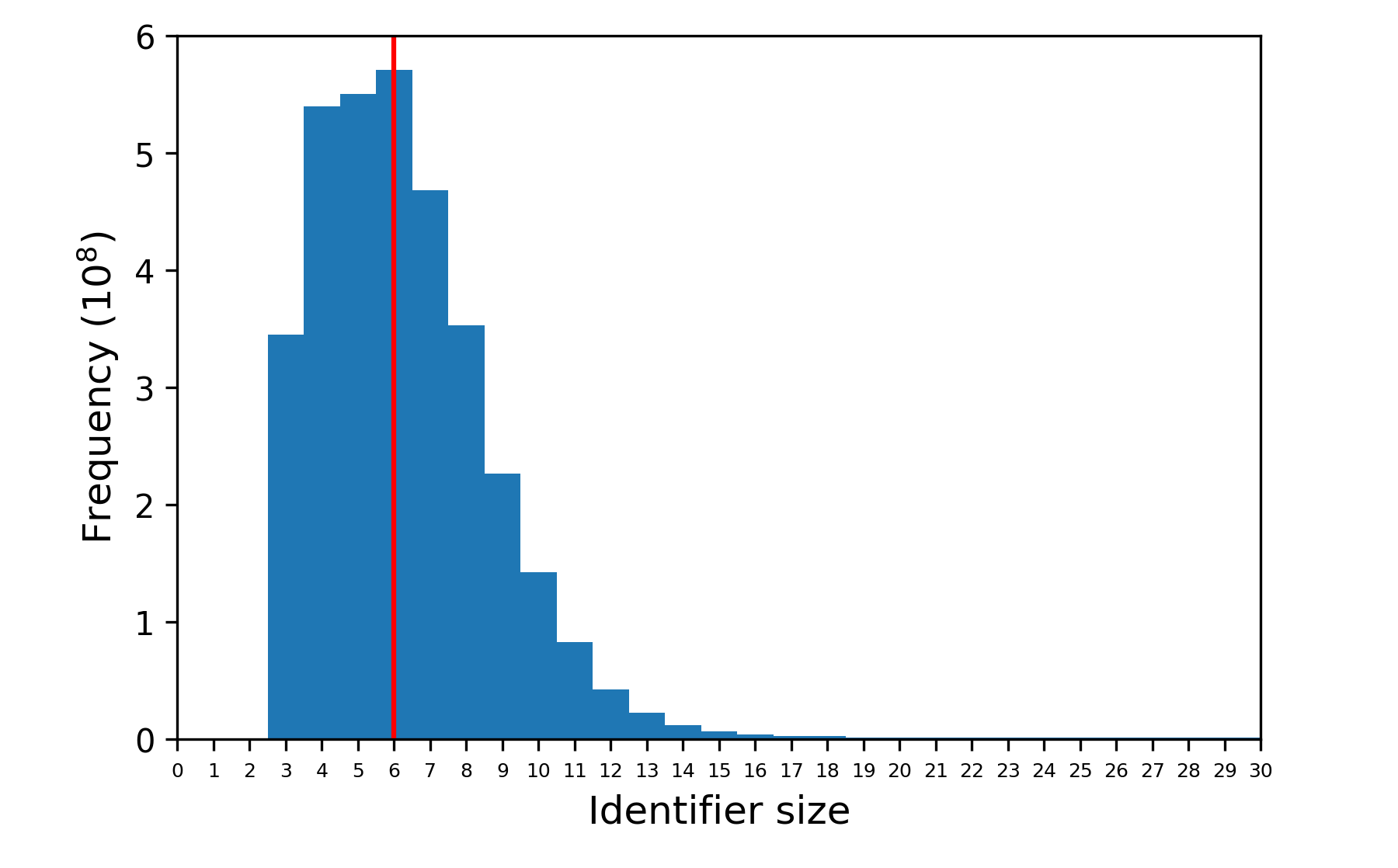}
\end{center}
\end{figure}

It can be seen that the most common name length is 6. Fig. \ref{stem_threshold} is the plot of the number of unique words in the dataset depending on the stemming threshold:

\begin{figure}
\caption{Influence of the stemming threshold on the vocabulary size}
\label{stem_threshold}
\begin{center}
\includegraphics[width=0.5\textwidth]{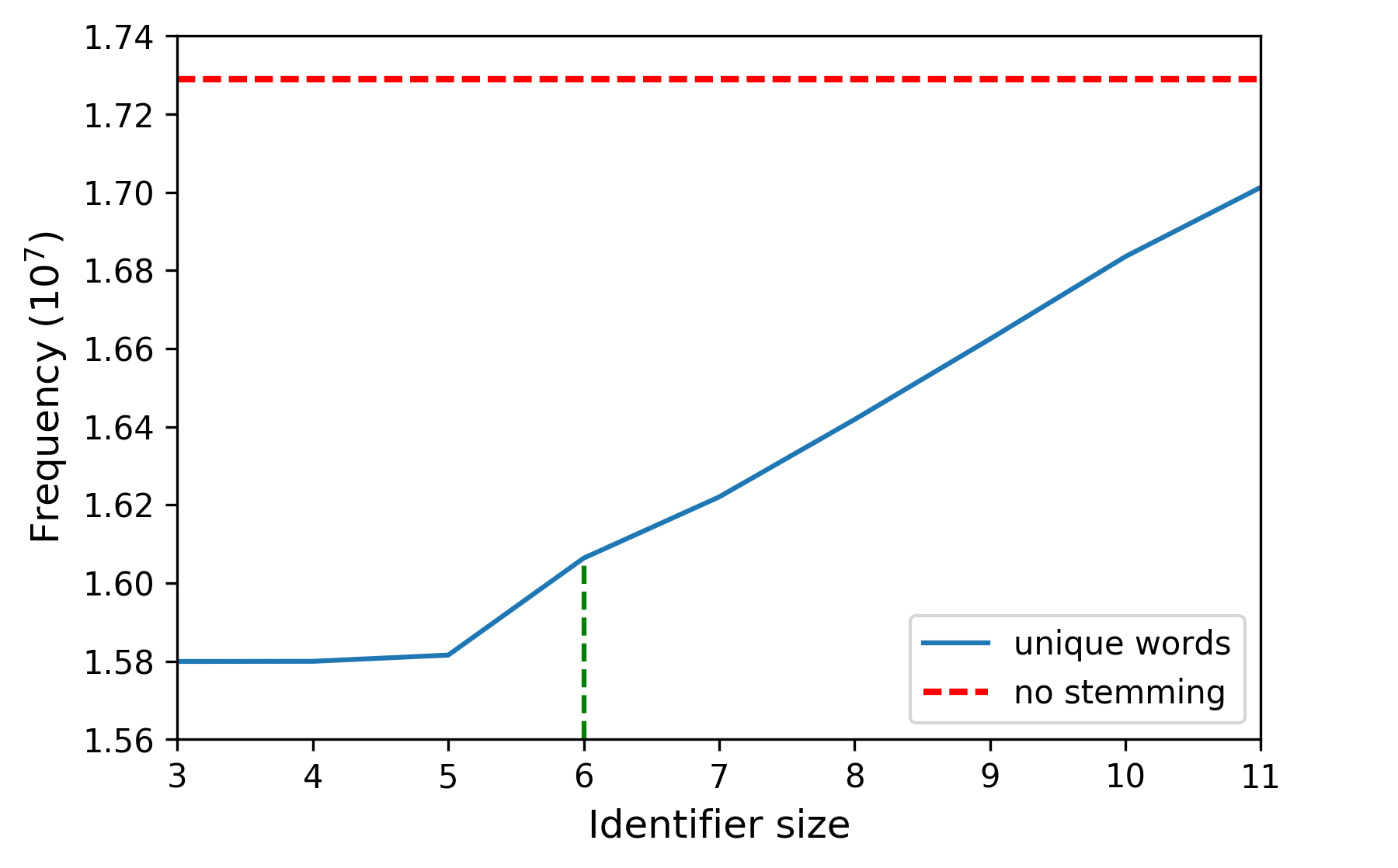}
\end{center}
\end{figure}

We observe the breaking point at length 5. The vocabulary size linearly grows starting with length 6. Having manually inspected several collisions on smaller thresholds, we came to the conclusion that 6 corresponds to the best trade-off between collisions and word normalization.

The Snowball algorithm was chosen based on the comparative study by Jivani \cite{Anjali_acomparative}. Stems not being real words are acceptable but it is critical to have minimum over-stemming since it increases the number of collisions. The total number of unique names are 16.06 million after stemming.

To be able to efficiently pre-process our data we used Apache Spark \cite{Spark} running on 64 4-core nodes which allowed us to process repositories in parallel in less than 1 day. 

However, before training a topic model one has to exclude near-duplicate repositories. In many cases GitHub users
include the source code of existing projects without preserving the commit history. For example, it is common for web sites, blogs and Linux-based firmwares. Those repositories contain very little original changes and may introduce frequency noise into the overall names distribution. This paper suggests the way to filter the described fuzzy duplicates based on the bag of words model built on the names in the source code.

\section{Filtering near-duplicate repositories} \label{fuzzy_clones}
There were more than 70 million GitHub repositories in October 2016 by our estimation. Approx. 18 million were not marked as forks. Nearly 800,000 repositories were de facto forks but not marked correspondingly by GitHub. That is, they had the same git commit history with colliding hashes. Such repositories may appear when a user pushes a cloned or imported repository under his or her own account without using the GitHub web interface to initiate a fork.

When we remove such hidden forks from the initial 18 million repositories, there still remain repositories which are highly similar. A duplicate repository is sometimes the result of a git push of an existing project with a small number of changes. For example, there are a large number of repositories with the Linux kernel which are ports to specific devices. In another case, repositories containing web sites were created using a cloned web engine and preserving the development history. Finally, a large number of \href{https://pages.github.com/}{github.io} repositories are the same. Such repositories contain much text content and few identifiers which are typically the same (HTML tags, CSS rules, etc.).

Filtering out such fuzzy forks speeds up the future training of the topic model and reduces the noise. As we obtained a bag-of-words for every repository, the naive approach would be to measure all pairwise similarities and find cliques. But first we need to define what is the similarity between bags-of-words.

\subsection{Weighted MinHash}
Suppose that we have two dictionaries - key-value mappings with unique keys and values indicating non-negative "weights" of the corresponding keys. We would like to introduce a similarity measure between them. The Jaccard Similarity between dictionaries $A=\{i: a_i\}, i\in I$ and $B=\{j: b_j\}$, $j\in J$ is defined as
\begin{equation}
J=\frac{\sum_ {k\in K}\limits \min(a_k, b_k)}{\sum_ {k\in K}\limits \max(a_k, b_k)}, K=I\cup J
\end{equation}
where $a_k = 0, k\notin I$ and $b_k = 0, k\notin J$. If the weights are
binary, this formula is equivalent to the common Jaccard Similarity definition.

The same way as MinHash is the algorithm to find similar sets in linear time, Weighted MinHash is the algorithm to find similar dictionaries in linear time. Weighted MinHash was introduced by Ioffe in \cite{Ioffe:2010:ICS:1933307.1934593}. We have chosen it in this paper because it is very efficient and allows execution on GPUs instead of large CPU clusters. The proposed algorithm depends on the parameter $K$ which adjusts the resulting hash length.

\begin{enumerate}[label=\arabic*.]
\item for $k$ in range($K$):
  \begin{enumerate}[label=1.\arabic*.]
  \item Sample $r_k, c_k \sim Gamma(2, 1)$ - Gamma distribution (their PDF is \mbox{$P(r)=re^{-r}$}), and
     $\beta_k \sim Uniform(0, 1)$.
  \item Compute
  \begin{align}
  t_ k &= \lfloor \frac{\ln S_k}{r_ k} + \beta_k\rfloor \\
  y_ k &= e^{r_k(t_k - \beta_k)} \\
  z_ k &= y_k e^{r_k} \\
  a_ k &= \frac{c_k}{z_k}
  \end{align}
  \end{enumerate}
\item Find $k^* = \arg\min_k a_k$ and return the samples $(k^*, t_{k^*})$.
\end{enumerate}

Thus given $K$ and supposing that the integers are 32-bit we obtain the hash
with size $8K$ bytes. Samples from $Gamma(2, 1)$ distribution can be efficiently calculated as $r = -\ln(u_1 u_2)$ where $u_1, u_2 \sim Uniform(0, 1)$ - uniform distribution between $0$ and $1$.

We developed the MinHashCUDA \cite{MHCUDA} library and Python native extension which is the implementation of Weighted MinHash algorithm for NVIDIA GPUs using CUDA \cite{Nickolls:2008:SPP:1365490.1365500}. There were several engineering challenges with that implementation which are unfortunately out of the scope of this paper. We were able to hash all 10 million repositories with hash size equal to 128 in less than 5 minutes using MinHashCUDA and 2 NVIDIA Titan X Pascal GPU cards.

\subsection{Locality Sensitive Hashing}
Having calculated all the hashes in the dataset, we can perform Locality Sensitive Hashing. We define several hash tables, each for it's own sub-hash which depends on the target level of false positives. Same elements will appear in the same bucket; union of the bucket sets across all the hash tables for a specific sample yields all the similar samples. Since our goal is to determine the sets of mutually similar samples, we should consider the set intersection instead.

We used the implementation of Weighted MinHash LSH from Datasketch \cite{datasketch}. It is designed after the corresponding algorithm in Mining of Massive Datasets \cite{Leskovec:2014:MMD:2787930}. LSH takes a single parameter - the target Weighted Jaccard Similarity value ("threshold"). MinHash LSH puts every repository in a number of separate hash tables which depend on the threshold and the hash size. We used the default threshold 0.9 in our experiments which ensures a low level of dissimilarity within a hash table bin.

\begin{figure}
\caption{LSH hash table's bin size distribution}
\label{bins_hist}
\begin{center}
\includegraphics[width=0.5\textwidth]{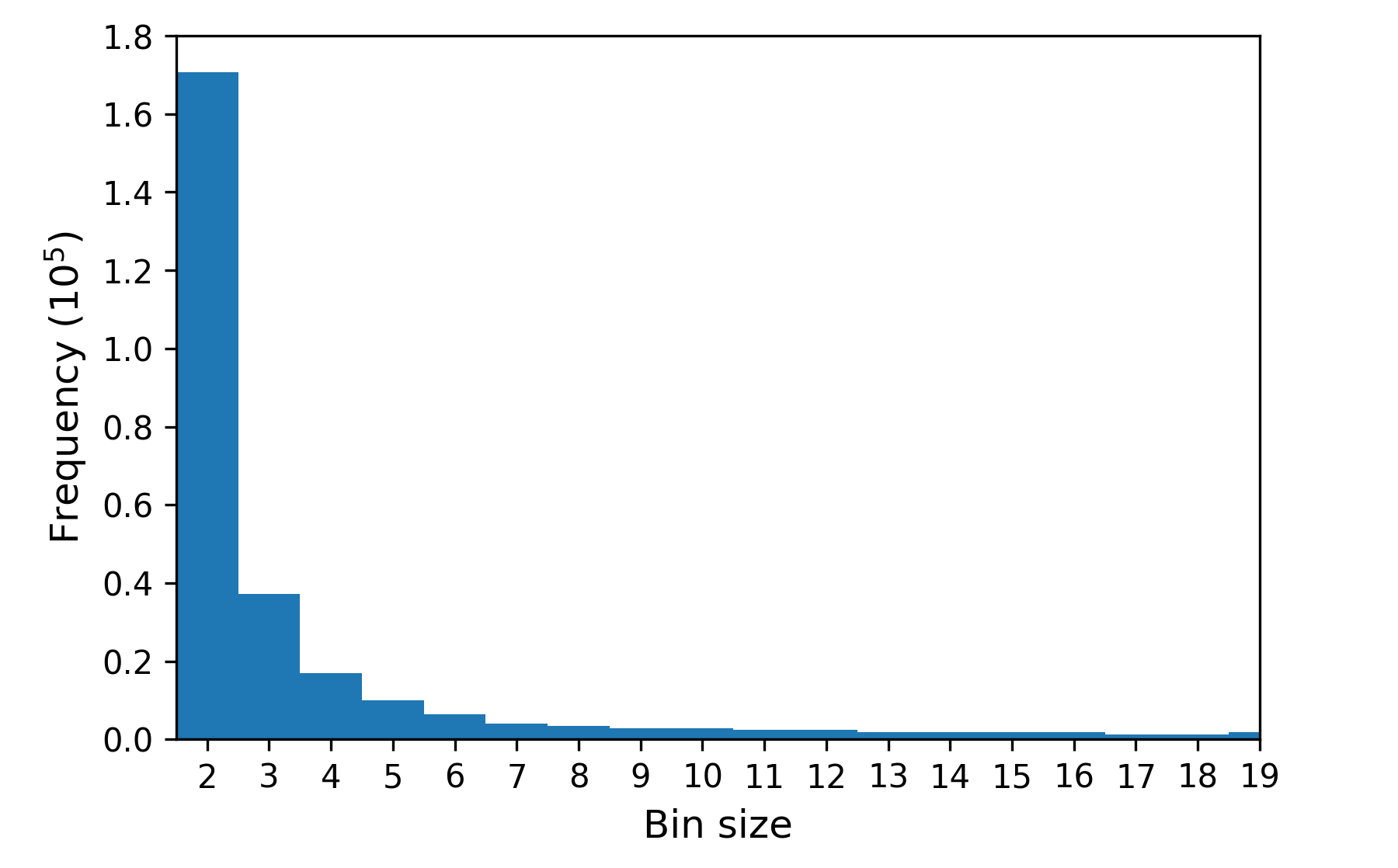}
\end{center}
\end{figure}

Algorithm \ref{lsh} describes the fuzzy duplicates detection pipeline. Step 6 discards less than 0.5\% of all the sets and aims at reducing the number of false positives. The bins size distribution after step 5 is depicted on Fig. \ref{bins_hist} - it is clearly seen that the majority of the bins has the size 2. Step 6 uses Weighted Jaccard similarity threshold 0.8 instead of 0.9 to be sensitive to evident outliers exclusively.

\begin{figure}
\caption{Fuzzy duplicates detection pipeline}
\label{lsh}
\begin{algorithmic}[1]
\State Calculate Weighted MinHash with hash size 128 for all the repositories.
\State Feed each hash to MinHash LSH with threshold 0.9 so that every repository appears in each of the 5 hash tables.
\State Filter out hash table bins with single entries.
\State For every repository, intersect the bins it appears in across all the hash tables. Cache the intersections, that is, if a repository appears in the same existing set, do not create the new one.
\State Filter out sets with a single element. The resulting number of unique repositories corresponds to "Filtered size" in Table \ref{hash_sizes}.
\State For every set with 2 items, calculate the precise Weighted Jaccard similarity value and filter out those with less than 0.8 (optional).
\State Return the resulting list of sets. The number of unique repositories corresponds to "Final size" in Table \ref{hash_sizes}.
\end{algorithmic}
\end{figure}

Table \ref{hash_sizes} reveals how different hash sizes influence on the resulting number of fuzzy clones:

\begin{table}
\caption{Influence of the Weighted MinHash size to the number of fuzzy clones}
\label{hash_sizes}
\centering
\begin{tabularx}{0.5\textwidth}{|l|l|l|l|X|}
\hline
\textbf{Hash size} & \textbf{Hash tables} & \textbf{Average bins} & \textbf{Filtered size} & \textbf{Final size} \\
\hline
64 & 3 & 272000 & 1,745,000 & 1,730,000 \\
128 & 5 & 263000 & 1,714,000 & 1,700,000 \\
160 & 6 & 261063 & 1,687,000 & 1,675,000 \\
192 & 7 & 258000 & 1,666,000 & 1,655,000 \\
\hline

\end{tabularx}
\end{table}

Algorithm \ref{lsh} results in approximately 467,000 sets of fuzzy duplicates with overall 1.7 million unique repositories. Each repository appears in two sets on average. The examples of fuzzy duplicates are listed in appendix \ref{fc_examples}.
The detection algorithm works especially well for static web sites which share the same JavaScript libraries.

\begin{figure}
\caption{Stemmed names frequency histogram}
\label{names_hist}
\begin{center}
\includegraphics[width=0.5\textwidth]{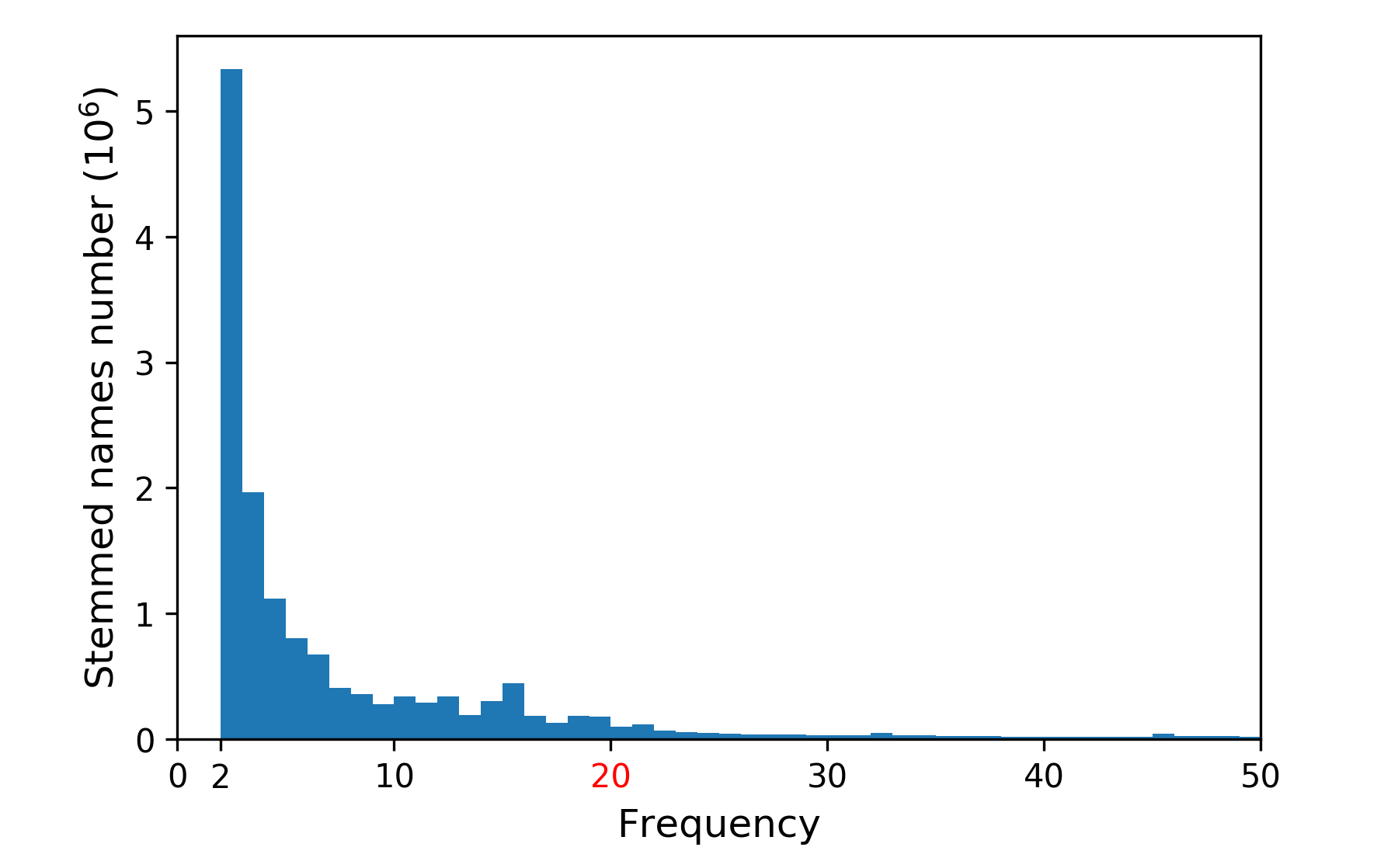}
\end{center}
\end{figure}

After the exclusion of the fuzzy duplicates, we finish dataset processing and pass over to training of the topic model. The total number of unique names has now reduced by 2.06 million to 14 million unique names. To build a meaningful dataset, names with occurrence of less than $T_f = 20$ were excluded from the final vocabulary. 20 was chosen on the frequency histogram shown on Fig. \ref{names_hist} since it is the drop-off point. 

\begin{figure}
\caption{Bag sizes after fuzzy duplicates filtering}
\label{bag_sizes}
\begin{center}
\includegraphics[width=0.5\textwidth]{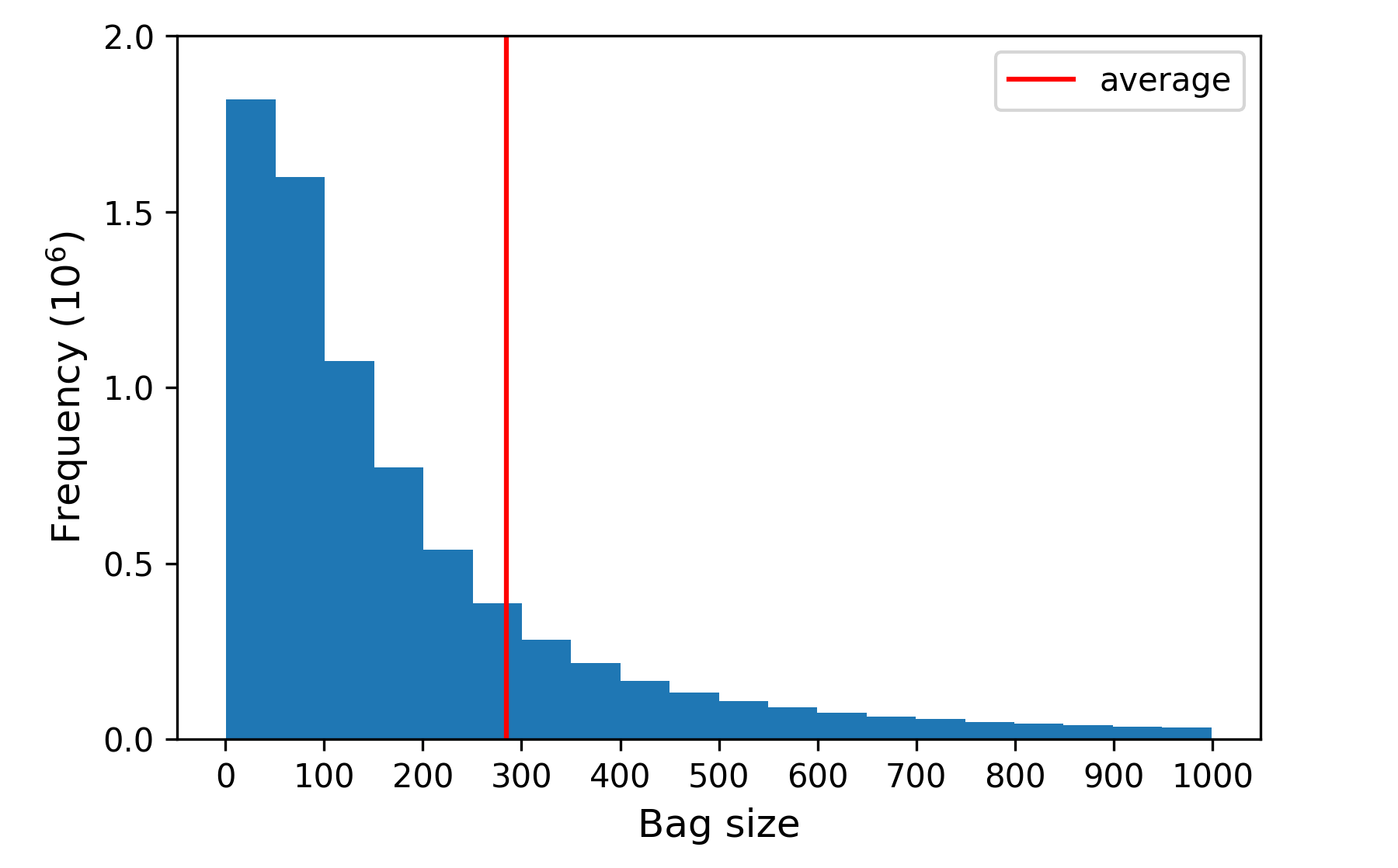}
\end{center}
\end{figure}

After this exclusion, there are now 2 million unique names, with an average size of a bag-of-words of 285 per repository and Fig. \ref{bag_sizes} displays the heavy-tailed bag size distribution.

\section{Training the ARTM topic model}
This section revises ARTMs and describes how the training of the topic model was performed. We have chosen ARTM instead of other topic modeling algorithms since it has the most efficient parallel CPU implementation in bigARTM according to our benchmarks.
\subsection{Additive Regularized Topic Model} 
Suppose that we have a topic probabilistic model of the collection of documents
$D$ which describes the occurrence of terms $w$ in document $d$ with topics $t$:
\begin{equation}
p(w|d) = \sum_{t\in T} p(w|t) p(t|d).
\end{equation}
Here $p(w|t)$ is the probability of the term $w$ to belong to the topic $t$,
$p(t|d)$ is the probability of the topic $t$ to belong to the document $d$,
thus the whole formula is just an expression of the total probability,
accepting the hypothesis of conditional independence: $p(w|d,t) = p(w|t)$.
Terms belong to the vocabulary $W$, topics are taken from the set $T$ which is simply the series of indices $[1, 2, \dots n_t]$.

We'd like to solve the problem of recovering $p(w|t)$ and $p(t|d)$ from
the given set of documents $\left\{d\in D: d = \left\{w_1 \dots w_{n_d}\right\}\right\}$.
We normally assume $\hat{p}(w|d) = \frac{n_{dw}}{n_d}$, $n_{dw}$
being the number of times term $w$ occurred in document $d$,
but this implies that all the terms are equally important which is not always true.
"Importance" here means some measure which negatively correlates with the overall frequency of the term.
Let us denote the recovered probabilities as $\hat{p}(w|t) = \phi_{wt}$ and $\hat{p}(t|d) = \theta_{td}$.
Thus our problem is the stochastic matrix decomposition which is not correctly stated:
\begin{equation}
\frac{n_{dw}}{n_d} \approx \Phi \cdot \Theta = (\Phi S)(S^{-1}\Theta) = \Phi' \cdot \Theta'.
\end{equation}

The stated problem can be solved by applying maximum likelihood estimation:
\begin{equation}
\sum_{d\in D}\sum_{w\in d}n_{dw}\ln \sum_{t}\phi_{wt} \theta_{td} \to \max_{\Phi,\Theta}
\end{equation}

upon the conditions

\begin{equation}
\phi_{wt} > 0; \sum_{w\in W}\phi_{wt} = 1;
\theta_{td} > 0; \sum_{t\in T}\theta_{td} = 1.
\end{equation}

The idea of ARTM is to naturally introduce regularization as one or several extra additive members:
\begin{equation}
\sum_{d\in D}\sum_{w\in d}n_{dw}\ln \sum_{t}\phi_{wt} \theta_{td} + R(\Phi,\Theta) \to \max_{\Phi,\Theta}.
\end{equation}

Since this is a simple summation, one can combine a series of regularizers in the same objective function. For example, it is possible to increase $\Phi$ and $\Theta$ sparsity or to make topics less correlated. Well-known LDA
 model \cite{Blei:2003:LDA:944919.944937} can be reproduced as ARTM too.

The variables $\Phi$ and $\Theta$ can be effectively calculated using the iterative expectation maximization algorithm \cite{College02theexpectation}.
Many ready to be used ARTM regularizers are already implemented in the BigARTM open source project \cite{BigARTM}.

\subsection{Training}
Vorontsov shows in \cite{Vorontsov2015} that ARTM is trained best if the regularizers are activated sequentially, with a lag relative to each other. For example, first EM iterations are performed without any regularizers at all and the model reaches target perplexity, then $\Phi$ and $\Theta$ sparsity regularizers are activated and the model optimizes for those new members in the objective function while not increasing the perplexity. Finally other advanced regularizers are appended and the model minimizes the corresponding members while leaving the old ones intact.

We apply only $\Phi$ and $\Theta$ sparsity regularizers in this paper. Further research is required to leverage others. We experimented with the training of ARTM on the source code identifiers from \ref{bag_of_words} and observed that the final perplexity and sparsity values do not change considerably on the wide range of adjustable meta-parameters. The best training meta-parameters are given in Table \ref{artm_meta}.
\begin{table}
\caption{Used ARTM meta-parameters}
\label{artm_meta}
\centering
\begin{tabular}{| l | l |}
\hline
\textbf{Parameter} & \textbf{Value} \\
\hline
Topics & 256 \\
Iterations without regularizers & 10 \\
Iterations with regularizers & 8 \\
$\Phi$ sparsity weight & 0.5 \\
$\Theta$ sparsity weight & 0.5 \\
\hline
\end{tabular}
\end{table}

\begin{figure}
\caption{ARTM convergence}
\label{topic_model}
\begin{center}
\includegraphics[width=0.5\textwidth]{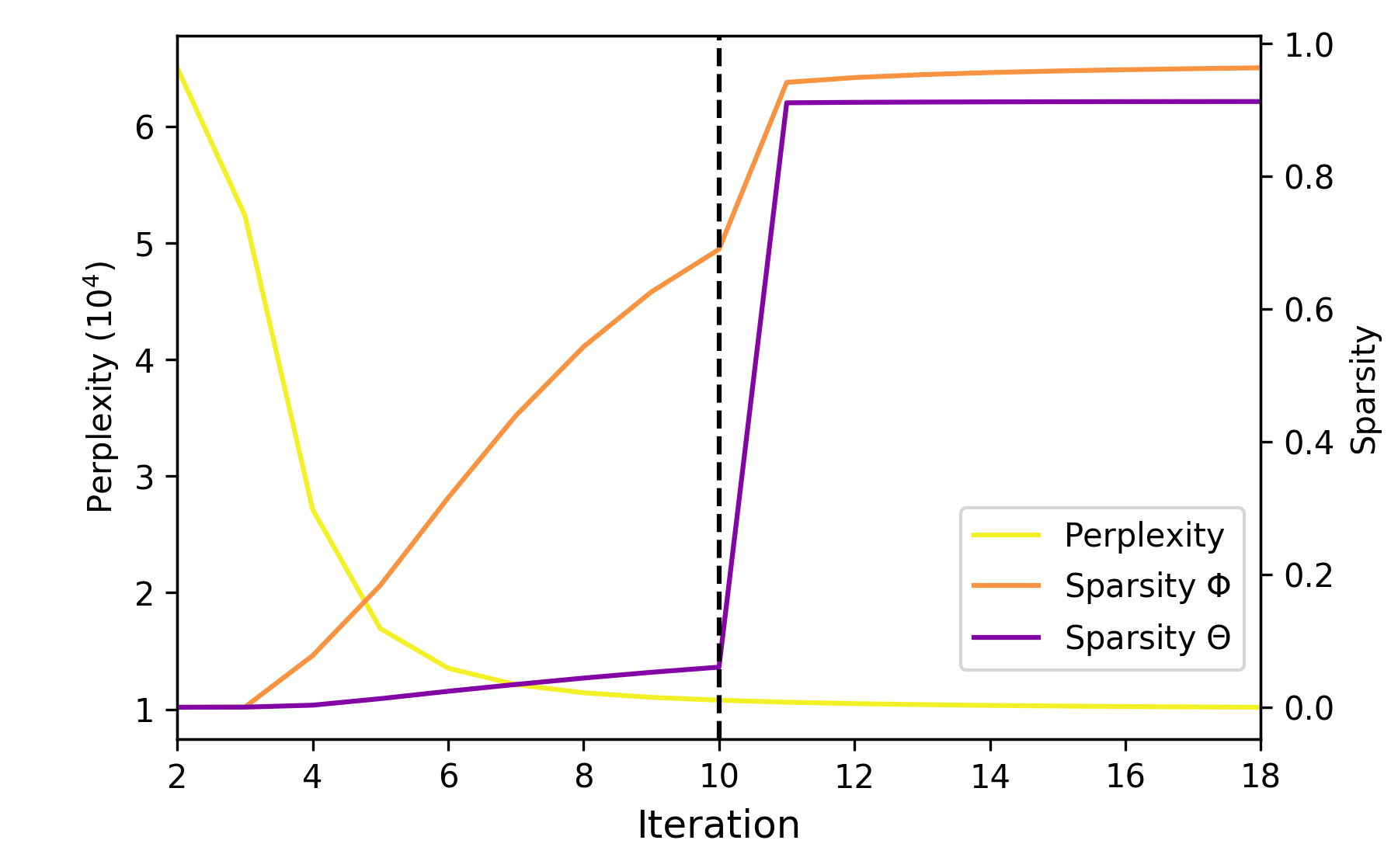}
\end{center}
\end{figure}

We chose 256 topics merely because it is time intensive to label them and 256 was the largest amount we could label. The traditional ways of determining the optimal number of topics using e.g. Elbow curves are not applicable to our data. We cannot consider topics as clusters since a typical repository corresponds to several topics and increasing the number of topics worsens the model's generalization and requires the dedicated topics decorrelation regularizer.
The overall number of iterations equals 18. The convergence plot is shown on Fig. \ref{topic_model}.

The achieved quality metric values are as given in Table \ref{artm_metrics}.

\begin{table}
\caption{Achieved ARTM metrics}
\label{artm_metrics}
\centering
\begin{tabular}{| l | l |}
\hline
\textbf{Metric} & \textbf{Value} \\
\hline
Perplexity & 10168 \\
$\Phi$ sparsity & 0.964 \\
$\Theta$ sparsity & 0.913 \\
\hline
\end{tabular}
\end{table}

On the average, a single iteration took 40 minutes to complete on our hardware. We used BigARTM in a Linux environment on 16-core (32 threads) Intel(R) Xeon(R) CPU E5-2620 v4 computer with 256 GB of RAM. BigARTM supports the parallel training and we set the number of workers to 30. The peak memory usage was approximately 32 GB.

It is possible to relax the hardware requirements and speed up the training if the model size is reduced. If we set the frequency threshold $T_f$ to a greater value, we can dramatically reduce the input data size with the risk of loosing the ability of the model to generalize.

We trained a reference LDA topic model to provide the baseline using the built-in LDA engine in BigARTM. 20 iterations resulted in perplexity 10336 and $\Phi$ and $\Theta$ sparsity 0 (fully dense matrices). It can be seen that the additional regularization not only made the model sparse but also yielded a better perplexity. We relate this observation with the fact that LDA assumes the topic distribution to have a sparse Dirichlet prior, which does not obligatory stand for our dataset.

\subsection{Converting the repositories to the topics space}
\begin{figure}
\caption{ARTM topic significance distribution}
\label{topics_sign}
\begin{center}
\includegraphics[width=0.5\textwidth]{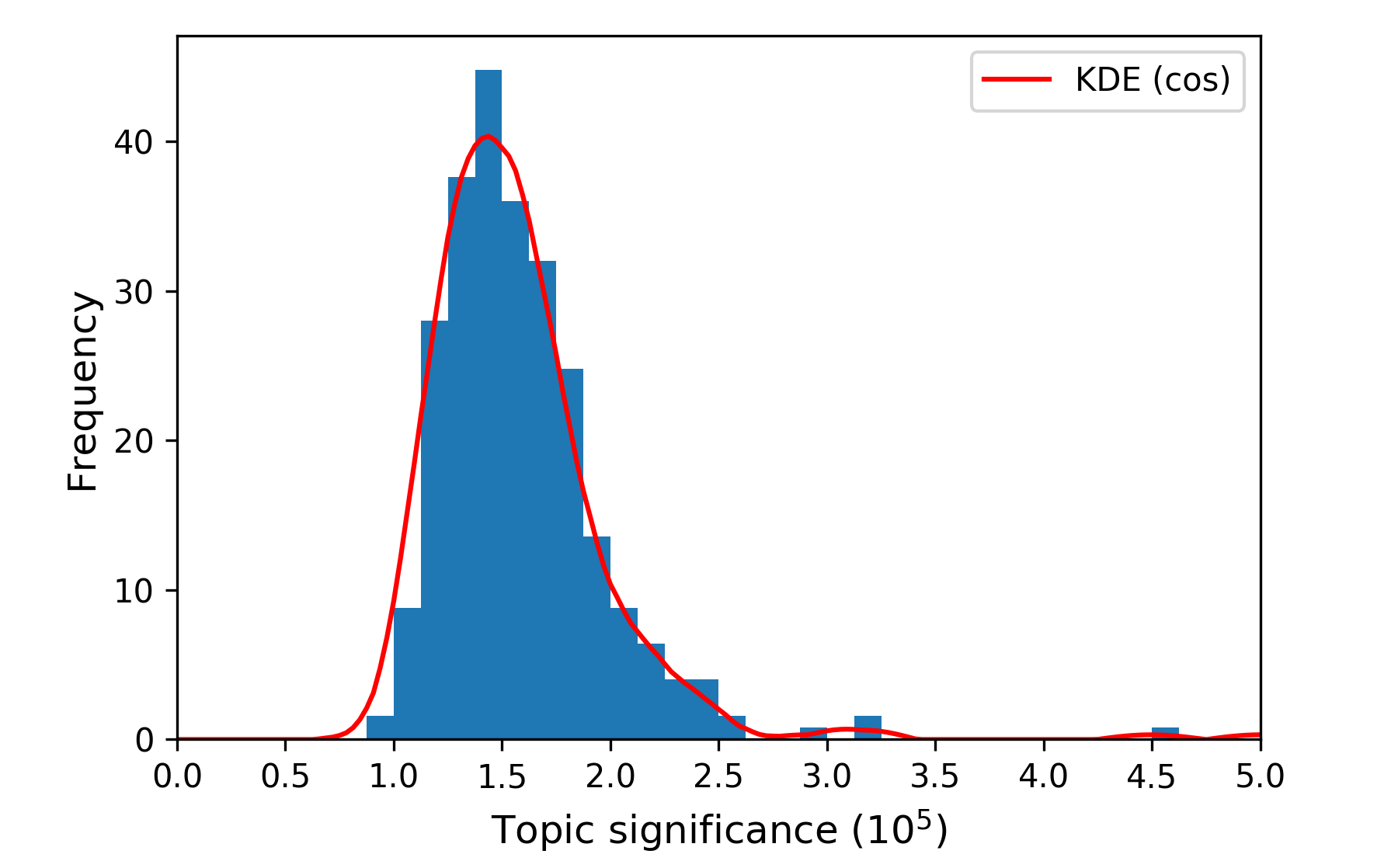}
\end{center}
\end{figure}

Let $R_t$ be the matrix of repositories in the topics space of size $R\times T$, $R_n$ be the sparse matrix representing the dataset of size $R\times N$ and $T_n$ be the matrix representing the trained topic model of size $N\times T$. We perform the matrix multiplication to get the repository embeddings:
\begin{equation}
R_t = R_n \times T_n.
\end{equation}
We further normalize each row of the matrix by $L2$ metric:
\begin{equation}
R_t^{normed} = rowwise\frac{R_t}{\lVert R_t\rVert_2}.
\end{equation}
The sum along every column of this matrix indicates the significance of each topic. Fig. \ref{topics_sign} shows the distribution of this measure.

\section{Topic modeling results} \label{results}
The employed topic model is unable to summarize the topics the same way humans do. It is possible to interpret some topics based on the most significant words, some based on relevant repositories, but many require manual supervision with the careful analysis of most relevant names and repositories. This supervision is labour intensive and the single topic normally takes up to 30 minutes to summarize with proper confidence. 256 topics required several man-days to complete the analysis.

After a careful analysis, we sorted the labelled topics into the following groups:

\begin{itemize}
\item \textbf{Concepts} (41) - general, broad and abstract. The most interesting group. It includes scientific terms, facts about the world and the society.
\item \textbf{Human languages} (10) - it appeared that one can determine programmer’s approximate native language looking at his code, thanks to the stem bias.
\item \textbf{Programming languages} (33) - not so interesting since this is the information we already have after \texttt{linguist} classification. Programming languages usually have a standard library of classes and functions which is imported/included into most of the programs, and the corresponding names are revealed by our topic modeling. Some topics are more narrow than a programming language.
\item \textbf{General IT} (72) - the topics which could appear in Concepts if had an expressive list of key words but do not. The repositories are associated by the unique set of names in the code without any special meaning.
\item \textbf{Technologies} (87) - devoted to some specific, potentially narrow technology or product. Often indicates an ecosystem or community around the technology.
\item \textbf{Games} (13) - related to video games. Includes specific gaming engines.
\end{itemize}

The complete topics list is in appendix \ref{topics}. The example topic labelled "Machine Learning, Data Science" is shown in appendix \ref{ml}.

\label{duplicateexpl}It can be observed that some topics are dual and need to be splitted. That duality is a sign that the number of topics should be bigger. At the same time, some topics appear twice and need to be de-correlated, e.g. using the "decorrelation" ARTM regularizer. Simple reduction or increase of the number of topics however do not solve those problems, we found it out while experimenting with 200 and 320 topics.

\section{Released datasets} \label{datasets}
We generated several datasets which were extracted from our internal 100 TB GitHub repository storage. We incorporated them on data.world \cite{data_world}, the recently emerged "GitHub for data scientists", each has the description, the origin note and the format definition. They are accessed at \href{https://data.world/vmarkovtsev}{data.world/vmarkovtsev}. Besides, the datasets are uploaded to Zenodo and have DOI. They are listed in Table \ref{datasets}.

\begin{table}[H]
\caption{Open datasets on data.world}
    \begin{tabularx}{0.5\textwidth}{| l | X |}
    \hline
    \textbf{Name abd DOI} & \textbf{Description} \\ \hline
    \pbox[t]{\textwidth}{source code names \\ \doi{10.5281/zenodo.284554}} & names extracted from 13,000,000 repositories (fuzzy clones excluded) considered in section \ref{bag_of_words} \\ \hline
    \pbox[t]{\textwidth}{452,000,000 commits \\ \doi{10.5281/zenodo.285467}} & metadata of all the commits in 16,000,000 repositories (fuzzy clones excluded)\\ \hline
    \pbox[t]{\textwidth}{keyword frequencies \\ \doi{10.5281/zenodo.285293}} & frequencies of programming language keywords (reserved tokens) across 16,000,000 repositories (fuzzy clones excluded)\\ \hline
    \pbox[t]{\textwidth}{readme files \\ \doi{10.5281/zenodo.285419}} & README files extracted from 16,000,000 repositories (fuzzy clones excluded) \\ \hline
    \pbox[t]{\textwidth}{duplicate repositories \\ \doi{10.5281/zenodo.285377}} & fuzzy clones which were considered in section \ref{fuzzy_clones} \\ \hline
    \end{tabularx}
\end{table}

\section{Conclusion and future work}
Topic modeling of GitHub repositories is an important step to understanding software development trends and open source communities. We built a repository processing pipeline and applied it to more than 18 million public repositories on GitHub. Using developed by us open source tool MinHashCUDA we were able to remove 1.6 million fuzzy duplicate repositories from the dataset. The preprocessed dataset with source code names as well as other datasets are open and the presented results can be reproduced. We trained ARTM on the resulting dataset and manually labelled 256 topics. The data processing and model training are possible to perform using a single GPU card and a moderately sized Apache Spark cluster. The topics covered a broad range of projects but there were repeating and dual ones. The chosen number of topics was enough for general exploration but not enough for the complete description of the dataset.

Future work may involve experimentation with clustering the repositories in the topic space and comparison with clusters based on dependency or social graphs \cite{Syed_onclusters}.

\appendices

\section{Parsed languages} \label{languages}
\begin{multicols}{4}
\begin{itemize}[label={},leftmargin=*]
\small
\item abap
\item abl
\item actionscript
\item ada
\item agda
\item ahk
\item alloy
\item antlr
\item apl
\item applescript
\item arduino
\item as3
\item aspectj
\item aspx-vb
\item autohotkey
\item autoit
\item awk
\item b3d
\item bash
\item batchfile
\item befunge
\item blitzbasic
\item blitzmax
\item bmax
\item boo
\item bplus
\item brainfuck
\item bro
\item bsdmake
\item c
\item c\#
\item c++
\item ceylon
\item cfc
\item cfm
\item chapel
\item chpl
\item cirru
\item clipper
\item clojure
\item cmake
\item cobol
\item coffeescript
\item coldfusion
\item common lisp
\item component pascal
\item console
\item coq
\item csharp
\item csound
\item cucumber
\item cuda
\item cython
\item d
\item dart
\item delphi
\item dosbatch
\item dylan
\item ec
\item ecl
\item eiffel
\item elisp
\item elixir
\item elm
\item emacs
\item erlang
\item factor
\item fancy
\item fantom
\item fish
\item fortran
\item foxpro
\item fsharp
\item gap
\item gas
\item genshi
\item gherkin
\item glsl
\item gnuplot
\item go
\item golo
\item gosu
\item groovy
\item haskell
\item haxe
\item hy
\item i7
\item idl
\item idris
\item igor
\item igorpro
\item inform 7
\item io
\item ioke
\item j
\item isabelle
\item jasmin
\item java
\item javascript
\item jsp
\item julia
\item kotlin
\item lasso
\item lassoscript
\item lean
\item lhaskell
\item lhs
\item limbo
\item lisp
\item literate agda
\item literate haskell
\item livescript
\item llvm
\item logos
\item logtalk
\item lsl
\item lua
\item make
\item mako
\item mathematica
\item matlab
\item mf
\item minid
\item mma
\item modelica
\item modula-2
\item monkey
\item moocode
\item moonscript
\item mupad
\item myghty
\item nasm
\item nemerle
\item nesc
\item newlisp
\item nimrod
\item nit
\item nix
\item nixos
\item nsis
\item numpy
\item obj-c
\item obj-c++
\item obj-j
\item objectpascal
\item ocaml
\item octave
\item ooc
\item opa
\item openedge
\item pan
\item pascal
\item pawn
\item perl
\item php
\item pike
\item plpgsql
\item posh
\item povray
\item powershell
\item progress
\item prolog
\item puppet
\item pyrex
\item python
\item qml
\item \makebox[\linewidth][l]{robotframework}
\end{itemize}
\end{multicols}
\begin{figure}[H]
\begin{multicols}{4}
\begin{itemize}[label={},leftmargin=*]
\item r
\item racket
\item ragel
\item rb
\item rebol
\item red
\item redcode
\item ruby
\item rust
\item sage
\item salt
\item scala
\item scheme
\item scilab
\item shell
\item shen
\item smali
\item smalltalk
\item smarty
\item sml
\item sourcepawn
\item splus
\item squeak
\item stan
\item standard ml
\item supercollider
\item swift
\item tcl
\item tcsh
\item thrift
\item typescript
\item vala
\item vb.net
\item verilog
\item vhdl
\item vim
\item winbatch
\item x10
\item xbase
\item xml+genshi
\item xml+kid
\item xquery
\item xslt
\item xtend
\item zephir
\end{itemize}
\end{multicols}
\end{figure}

\section{Examples of fuzzy duplicate repositories} \label{fc_examples}
\subsection{Linux kernel}
\begin{itemize}
\item 1406/linux-0.11
\item yi5971/linux-0.11
\item love520134/linux-0.11
\item wfirewood/source-linux-0.11
\item sunrunning/linux-0.11
\item Aaron123/linux-0.11
\item junjee/linux-0.11
\item pengdonglin137/linux-0.11
\item yakantosat/linux-0.11
\end{itemize}
\subsection{Tutorials}
\begin{itemize}
\item dcarbajosa/linuxacademy-chef
\item jachinh/linuxacademy-chef
\item flarotag/linuxacademy-chef
\item qhawk/linuxacademy-chef
\item paul-e-allen/linuxacademy-chef
\end{itemize}
\subsection{Web applications 1}
\begin{itemize}
\item choysama/my-django-first-blog
\item mihuie/django-first
\item PubMahesh/my-first-django-app
\item nickmalhotra/first-django-blog
\item Jmeggesto/MyFirstDjango
\item atlwendy/django-first
\item susancodes/first-django-app
\item quipper7/DjangoFirstProject
\item phidang/first-django-blog
\end{itemize}
\subsection{Web applications 2}
\begin{itemize}
\item iggitye/omrails
\item ilrobinson81/omrails
\item OCushman/omrails
\item hambini/One-Month-Rails
\item Ben2pop/omrails
\item chrislorusso/omrails
\item arjunurs/omrails
\item crazystingray/omrails
\item scorcoran33/omrails
\item Joelf001/Omrails
\end{itemize}

\newcounter{topics}
\begin{figure}[H]
\section{Complete list of labelled topics} \label{topics}
\subsection{Concepts}
\begin{multicols}{2}
\begin{enumerate}
    \item 2D geometry
    \item 3D geometry
    \item Arithmetic
    \item Audio
    \item Bitcoin
    \item Card Games
    \item Chess; Hadoop \hyperref[dual]{\#}
    \item Classical mechanics (physics)
    \item Color Manipulation/Generation
    \item Commerce, ordering
    \item Computational Physics 
    \item Date and time
    \item Design patterns; HTML parsing
    \item Email \hyperref[duplicate]{*}
    \item Email \hyperref[duplicate]{*}
    \item Enumerators, Mathematical Expressions
    \item Finance and trading
    \item Food (eg. pizza, cheese, beverage), Calculator
    \item Genomics
    \item Geolocalization, Maps
    \item Graphs
    \item Hexademical numbers
    \item Human
    \item Identifiers
    \item Language names; JavaFX \hyperref[dual]{\#}
    \item Linear Algebra; Optimization
    \item Machine Learning, Data Science
    \item My
    \item Parsing
    \item Particle physics
    \item Person Names (American)
    \item Personal Information
    \item Photography, Flickr
    \item Places, transportation, travel
    \item Publishing; Flask \hyperref[dual]{\#}
    \item Space and solar system
    \item Sun and moon
    \item Trade
    \item Trees, Binary Trees
    \item Video; movies
    \item Word Term
    \setcounter{topics}{\value{enumi}}
 \end{enumerate}
 \end{multicols}
 \subsection{Human languages}
 \begin{multicols}{2}
 \begin{enumerate}
    \setcounter{enumi}{\value{topics}}
    \item Chinese
    \item Dutch
    \item French \hyperref[duplicate]{*}
    \item French \hyperref[duplicate]{*}
    \item German
    \item Portuguese \hyperref[duplicate]{*}
    \item Portuguese \hyperref[duplicate]{*}
    \item Spanish \hyperref[duplicate]{*}
    \item Spanish \hyperref[duplicate]{*}
    \item Vietnamese
    \setcounter{topics}{\value{enumi}}
 \end{enumerate}
 \end{multicols}
 \subsection{Programming languages}
 \begin{multicols}{2}
 \begin{enumerate}
    \setcounter{enumi}{\value{topics}}
    \item Assembler
    \item Autoconf
    \item Clojure
    \item ColdFusion \hyperref[duplicate]{*}
    \item ColdFusion \hyperref[duplicate]{*}
    \item Common LISP
    \item Emacs LISP
    \item Emulated assembly
    \item Go
    \item HTML
    \item Human education system
    \item Java AST and bytecode
    \item libc
    \item Low-level PHP
    \item Lua \hyperref[duplicate]{*}
    \item Lua \hyperref[duplicate]{*}
    \item Makefiles
    \item \makebox[\linewidth][l]{Mathematics: proofs, sets}
    \item Matlab
    \item Object Pascal
    \item Objective-C
    \item Perl
    \item Python
    \item Python, ctypes
    \item Ruby
    \item Ruby with language extensions
    \item SQL
    \item String Manipulation in C
    \setcounter{topics}{\value{enumi}}
 \end{enumerate}
 \end{multicols}
 \end{figure}
 \begin{figure}[H]
 \begin{multicols}{2}
 \begin{enumerate}
    \setcounter{enumi}{\value{topics}}
    \item Verilog/VHDL
    \item Work, money, employment, driving, living
    \item x86 Assembler \hyperref[duplicate]{*}
    \item x86 Assembler \hyperref[duplicate]{*}
    \item XPCOM
    \setcounter{topics}{\value{enumi}}
 \end{enumerate}
 \end{multicols}
 \subsection{General IT}
 \begin{multicols}{2}
 \begin{enumerate}
    \setcounter{enumi}{\value{topics}}
    \item 3-char identifiers
    \item Advertising (Facebook, Ad Engines, Ad Blockers, AdMob)
    \item Animation
    \item Antispam; PHP forums
    \item Antivirus; database access \hyperref[dual]{\#}
    \item Barcodes; browser engines \hyperref[dual]{\#}
    \item Charting
    \item Chat; messaging
    \item Chinese web
    \item Code analysis and generation
    \item Computer memory and interfaces
    \item Console, terminal, COM
    \item CPU and kernel
    \item Cryptography
    \item Date and time picker
    \item DB Sharding, MongoDB sharding
    \item Design patterns; formal architecture
    \item DevOps
    \item Drawing \hyperref[duplicate]{*}
    \item Drawing \hyperref[duplicate]{*}
    \item Forms (UI)
    \item Glyphs; X11 and FreeType \hyperref[dual]{\#}
    \item Grids and tables
    \item HTTP auth
    \item iBeacons
    \item Image Manipulation
    \item Image processing
    \item Intel SIMD, Linear Algebra \hyperref[dual]{\#}
    \item IO operations
    \item Javascript selectors
    \item JPEG and PNG
    \item Media Players
    \item Metaprogramming
    \item Modern JS frontend (Bower, Grunt, Yeoman)
    \item Names starting with “m”
    \item Networking
    \item OAuth; major web services \hyperref[dual]{\#}
    \item Observer design pattern
    \item Online education; Moodle
    \item OpenGL \hyperref[duplicate]{*}
    \item Parsers and compilers
    \item Plotting
    \item Pointers
    \item POSIX Shell; VCS \hyperref[dual]{\#}
    \item Promises and deferred execution; Angular \hyperref[dual]{\#}
    \item Proof of concept
    \item RDF and SGML parsing
    \item Request and Response
    \item Requirements and dependencies
    \item Sensors; DIY devices
    \item Sockets C API
    \item Sockets, Networking
    \item Sorting and searching
    \item SQL database
    \item SQL DB, XML in PHP projects
    \item SSL
    \item Strings
    \item Testing with mocks
    \item Text editor UI
    \item Threads and concurrency
    \item Typing suggestions and dropdowns
    \item UI
    \item Video player
    \item VoIP
    \item Web Media, Arch Packages \hyperref[dual]{\#}
    \item Web posts
    \item Web testing; crawling
    \item Web UI
    \item Wireless
    \item Working with buffers
    \item XML (SAX, XSL)
    \item XMPP
    \item .NET
    \item Android Apps
    \item Android UI
    \item Apache Libraries for BigData
    \item Apache Thrift
    \item Arduino, AVR
    \item ASP.NET \hyperref[duplicate]{*}
    \item ASP.NET \hyperref[duplicate]{*}
    \setcounter{topics}{\value{enumi}}
 \end{enumerate}
 \end{multicols}
 \end{figure}
 \begin{figure}[H]
 \subsection{Technologies}
 \begin{multicols}{2}
 \begin{enumerate}
    \setcounter{enumi}{\value{topics}}
    \item Backbone.js
    \item Chardet (Python)
    \item Cocos2D
    \item \makebox[\linewidth][l]{Comp. vision; OpenCV}
    \item Cordova
    \item CPython
    \item Crumbs; cake(PHP)
    \item cURL
    \item DirectDraw
    \item DirectX
    \item Django Web Apps, CMS
    \item Drupal
    \item Eclipse SWT
    \item Emacs configs
    \item Emoji and Dojo \hyperref[dual]{\#}
    \item Facebook; Parse SDK \hyperref[dual]{\#}
    \item ffmpeg
    \item FLTK
    \item Fonts
    \item FPGA, Verilog
    \item FreeRTOS (Embedded)
    \item Glib
    \item Ionic framework, Cordova
    \item iOS Networking
    \item iOS Objective-C API
    \item iOS UI
    \item Jasmine tests, JS exercises, exercism \hyperref[dual]{\#}
    \item Java GUI
    \item Java Native Interface
    \item Java web servers
    \item Javascript AJAX, Javascript DOM manipulation
    \item Joomla
    \item JQuery
    \item jQuery Grid
    \item Lex, Yacc compiler
    \item libav / ffmpeg
    \item Linear algebra libraries
    \item Linux Kernel, Linux Wireless
    \item Lodash
    \item MFC Desktop Applications 
    \item Minecraft mods
    \item Monads
    \item OpenCL
    \item OpenGL \hyperref[duplicate]{*}
    \item PHP sites written by non-native English people
    \item PIC32
    \item Portable Document Format
    \item Puppet
    \item Pusher.com Apps
    \item Python packaging
    \item Python scientific stack
    \item Python scrapers
    \item Qt \hyperref[duplicate]{*}
    \item Qt \hyperref[duplicate]{*}
    \item React
    \item ROS (Robot Operating System)
    \item Ruby On Rails Apps 
    \item SaltStack
    \item Shockwave Flash
    \item Spreadsheets (Excel)
    \item Spreadsheets with PHP
    \item SQLite
    \item STL, Boost
    \item STM32
    \item Sublime Extensions
    \item Symphony, Doctrine; NLP \hyperref[dual]{\#}
    \item U-boot
    \item Vim Extensions
    \item Visual Basic, MSSQL
    \item Web scraping
    \item WinAPI
    \item Wordpress \hyperref[duplicate]{*}
    \item Wordpress \hyperref[duplicate]{*}
    \item Wordpress-like frontend
    \item Working with PDF in PHP
    \item wxWidgets
    \item Zend framework
    \item zlib \hyperref[duplicate]{*}
    \item zlib \hyperref[duplicate]{*}
    \setcounter{topics}{\value{enumi}}
 \end{enumerate}
 \end{multicols}
 \subsection{Games}
 \begin{multicols}{2}
 \begin{enumerate}
    \setcounter{enumi}{\value{topics}}
    \item 3D graphics and Unity
    \item Fantasy Creatures
    \setcounter{topics}{\value{enumi}}
 \end{enumerate}
 \end{multicols}
 \end{figure}
 \begin{figure}[H]
 \small\label{duplicate}* Repeating topic with different key words, see section \ref{duplicateexpl}.
 
 \label{dual}\# Dual topic, see section \ref{duplicateexpl}.
 \end{figure}
 \begin{figure}[H]
 \begin{multicols}{2}
 \begin{enumerate}
    \setcounter{enumi}{\value{topics}}
    \item Games
    \item Hello World, Games
    \item Minecraft
    \item MMORPG
    \item Pokemon
    \item Puzzle games
    \item RPG, fantasy
    \item Shooters (SDL)
    \item Unity Engine
    \item Unity3D Games
    \item Web Games
\end{enumerate}
\end{multicols}
\end{figure}

\section{Key words and repositories belonging to topic \#27 (Machine Learning, Data Science)} \label{ml}
\begin{figure}[H]
\centering \small
\begin{tabularx}{0.5\textwidth}{| l | l || l | X |}
\hline
\textbf{Rank} & \textbf{Word} & \textbf{Rank} & \textbf{Repository} \\
\hline
0.313115 & plot & 1.000000 & jingxia/kaggle\_yelp \\
0.303456 & numpy & 0.999998 & Carreau/spicy \\
0.273759 & plt & 0.999962 & zck17388/test \\
0.187565 & figur & 0.999719 & jonathanekstrand/Python\_... \\
0.181307 & zeros & 0.999658 & skendrew/astroScanr \\
0.169696 & matplotlib & 0.999543 & southstarj/YCSim \\
0.166166 & dtype & 0.999430 & parteekDhream/statthermo... \\
0.165236 & fig & 0.999430 & axellundholm/FMN050 \\
0.159658 & ylabel & 0.999361 & soviet1977/PriceList \\
0.153094 & xlabel & 0.999354 & connormarrs/3D-Rocket-... \\
0.146327 & subplot & 0.999282 & Holiver/matplot \\
0.144736 & shape & 0.999103 & wetlife/networkx \\
0.132792 & pyplot & 0.999034 & JingshiPeter/CS373 \\
0.124264 & scipy & 0.998969 & marialeon/los4mas2 \\
0.120666 & axis & 0.998385 & acnz/Project \\
0.110212 & arang & 0.998138 & khintz/GFSprob \\
0.110049 & mean & 0.998123 & claralusan/test\_aug\_18 \\
0.096037 & reshap & 0.997822 & amcleod5/PythonPrograms \\
0.093182 & range & 0.997662 & ericqh/deeplearning \\
0.084059 & ylim & 0.997567 & laserson/stitcher \\
0.082812 & linspac & 0.996786 & hs-jiang/MCA\_python \\
0.081260 & savefig & 0.996327 & DianaSplit/Tracking \\
0.080978 & xlim & 0.995153 & SivaGabbi/Ipython-Noteb... \\
0.080325 & axes & 0.994776 & prov-suite/prov-sty \\
0.077891 & legend & 0.992801 & bmoerker/EffectSizeEstim... \\
0.076858 & bins & 0.992558 & natalink/machine\_learning \\
0.076140 & panda & 0.992324 & olehermanse/INF1411-El... \\
0.076043 & astyp & 0.991026 & fonnesbeck/scipy2014\_tut... \\
0.075235 & pylab & 0.990514 & fablab-paderborn/device-... \\
0.073265 & ones & 0.989586 & mqchau/dataAnalysis \\
0.072214 & xrang & 0.988722 & acemaster/Image-Process... \\
0.072196 & len & 0.988514 & henryoswald/Sin-Plot \\
0.069818 & float & 0.987874 & npinto/virtualenv-bootstrap \\
0.065453 & linewidth & 0.987039 & ipashchenko/test\_datajoy \\
0.065453 & linalg & 0.986802 & Ryou-Watanabe/practice \\
0.065322 & norm & 0.986272 & mirthbottle/datascience-... \\
0.064042 & hist & 0.986039 & hglabska/doktorat \\
0.062975 & label & 0.985837 & parejkoj/yaledemo \\
0.061608 & sum & 0.985246 & grajasumant/python \\
0.060443 & cmap & 0.985173 & aaronspring/Scientific-Py... \\
0.059155 & scatter & 0.985043 & asesana/plots \\
0.058877 & fontsiz & 0.984838 & Sojojo83/SJFirstRepository \\
0.057343 & self & 0.983808 & Metres/MetresAndThtu... \\
0.057328 & none & 0.983393 & e-champenois/pySurf \\
0.056908 & true & 0.983170 & pawel-kw/dexy-latex-ex... \\
0.056292 & xtick & 0.983074 & keiikegami/envelopetheorem \\
0.051978 & figsiz & 0.982967 & msansa/test \\
0.051359 & sigma & 0.982904 & qdonnellan/\mbox{spyder-examples} \\
0.050785 & ndarray & 0.981725 & qiuwch/PythonNotebook... \\
0.050586 & sqrt & 0.981156 & rescolo/getdaa \\
\hline
\end{tabularx}
\end{figure}

\clearpage
\pagebreak
\bibliography{topic_modeling}
\bibliographystyle{ieeetr}

\end{document}